\documentclass[onecolumn,aps,prd,nofootinbib,10pt]{revtex4-2}

 \usepackage{color,graphicx,epsfig}
 \usepackage{amsmath,amssymb,dsfont,epsfig,graphicx,xcolor}
 \usepackage{ifpdf}
 \usepackage{amsmath}
 \usepackage{bm}
 \usepackage{color}
 \usepackage[english]{babel}
 \usepackage{graphicx}%
 \usepackage{amsfonts}%
 \usepackage{amssymb}
 \usepackage{braket}
 \usepackage{epstopdf}
\usepackage[utf8]{inputenc}

\usepackage{siunitx}

\definecolor{nicered}{rgb}{0.7,0.1,0.1}
\definecolor{nicegreen}{rgb}{0.1,0.5,0.1}
\definecolor{violet}{rgb}{0.7,0.3,0.3}
\usepackage[colorlinks,citecolor= nicegreen,linkcolor= nicered]{hyperref}
\usepackage[capitalize,nameinlink]{cleveref}

\newcommand{\nc}{\newcommand}
\nc{\non}{\nonumber}
\nc{\hc}{\hbox {H.c.}}
\nc{\noi}{\noindent}
\nc{\barx}{\bar{x}}
\nc{\pbarn}{\;\hbox {pb}}
\nc{\fbarn}{\;\hbox {fb}}

\nc{\hsp}{\hspace{0.5cm}}
\nc{\lsp}{\hspace{1cm}}
\nc{\Lsp}{\hspace{2cm}}
\nc{\LLsp}{\lsp\lsp}
\nc{\lra}{\longrightarrow}
\nc{\p}{\prime}
\nc{\sgn}{\text{sgn}}
\nc{\ph}{\varphi}
\nc{\op}{{\cal O}}
\nc{\eq}{\text{Eq.~}}
\nc{\beq}{\begin{equation}}  \nc{\eeq}{\end{equation}}
\nc{\bea}{\begin{eqnarray}}  \nc{\eea}{\end{eqnarray}}
\nc{\baa}{\begin{array}}     \nc{\eaa}{\end{array}}
\nc{\bit}{\begin{itemize}}   \nc{\eit}{\end{itemize}}
\nc{\ben}{\begin{enumerate}} \nc{\een}{\end{enumerate}}
\nc{\bce}{\begin{center}}    \nc{\ece}{\end{center}}
\nc{\bpm}{\begin{pmatrix}}   \nc{\epm}{\end{pmatrix}}
\nc{\bvt}{\begin{verbatim}}  \nc{\evt}{\end{verbatim}}

\begin{document}

\def\LjubljanaFMF{Faculty of Mathematics and Physics, University of Ljubljana,
 Jadranska 19, 1000 Ljubljana, Slovenia }
\def\LjubljanaIJS{Jo\v zef Stefan Institute, Jamova 39, 1000 Ljubljana, Slovenia}
\def\Orsay{Universit\'e Paris-Saclay, CNRS/IN2P3, IJCLab, 91405 Orsay, France}

\title{Testing Lepton Flavour Universality in \texorpdfstring{$\Upsilon (4S)$}{Y(4S)} Decays}

\author{S\'ebastien Descotes-Genon}
\email[Electronic address:]{sebastien.descotes-genon@ijclab.in2p3.fr}
\affiliation{\Orsay}

\author{Svjetlana Fajfer}
\email[Electronic address:]{svjetlana.fajfer@ijs.si}
\affiliation{\LjubljanaIJS}
\affiliation{\LjubljanaFMF}

\author{Jernej~F.~Kamenik}
\email[Electronic address:]{jernej.kamenik@cern.ch}
\affiliation{\LjubljanaIJS}
\affiliation{\LjubljanaFMF}

\author{Mart\'in Novoa-Brunet}
\email[Electronic address:]{martin.novoa@ijclab.in2p3.fr}
\affiliation{\Orsay}

\date{\today}
\begin{abstract}

  {We propose a novel method to probe the persistent hints of Lepton Flavour Universality violation observed in semileptonic $B$ decays. Relying on the specific properties of the Belle II experiment, it consists in comparing the inclusive rates of
$\Upsilon(4S) \to e^\pm \mu^\mp X$,  $\Upsilon(4S) \to \mu^\pm \tau_{\rm had}^\mp X$ and $\Upsilon(4S) \to e^\pm \tau_{\rm had}^\mp X$. We show that such a measurement can be directly related to the ratio $R(X)_{\tau\ell} \equiv \Gamma (b\to X \tau \nu ) / \Gamma (b \to X \ell \nu)$ ($\ell=e$ or $\mu$), once appropriate experimental cuts are applied to suppress the effects of neutral $B$ mixing and leptons emitted through charm or tau decays. Such a measurement would thus constitute an additional and potentially competitive probe of Lepton Flavour Universality in $b\to c\ell\nu$ transitions, complementary to existing exclusive measurements, accessible in the Belle II environment.
}

\end{abstract}

\maketitle


\section{Introduction}\label{intro}
The universality of electroweak interactions of leptons is a direct consequence of the gauge structure of the Standard Model (SM). Within the SM, lepton flavour universality (LFU) is violated solely due to the different masses of the charged leptons.
In the past several years a number of experiments observed the violation of LFU in the $b\to c \tau \nu_\tau$ decays, represented by the ratios $R_{D^{(\ast)}} =\mathcal B( B\to D^{(*)} \tau \nu_\tau) /\mathcal B( B\to D^{(*)} l \nu_l) $ \cite{Amhis:2019ckw,Lees:2012xj,Lees:2013uzd,Huschle:2015rga,Aaij:2015yra,Hirose:2016wfn,Hirose:2017dxl,Aaij:2017uff,Aaij:2017deq,Abdesselam:2019dgh}.
Heavy Flavour Averaging Group (HFLAV)~\cite{Amhis:2019ckw}  currently reports $R_{D^\ast}= 0.295\pm 0.011\pm 0.008$ and $R_{D}= 0.340\pm 0.027 \pm0.013$, which differ from the SM predictions $R_{D^\ast}^{SM}= 0.258\pm 0.0005$ and $R_{D}^{SM}= 0.299\pm 0.003$ on the level of $\sim 3 \,\sigma$.
A related  $1.8 \sigma$ tension has also been reported in the $B_c\to J/\psi \tau \nu$ decay branching ratio~\cite{Aaij:2017tyk}. 

In light of these intriguing results, several associated tests of LFU have been proposed in processes involving third generation quarks and leptons, including the rare $B$ meson decays $B_c \to \tau\nu$~\cite{Becirevic:2016yqi,Alonso:2016oyd},  $B \to K^{(*)} \nu\bar\nu$~\cite{Descotes-Genon:2020buf}, $B\to K^{(*)} \tau^+ \tau^-$~\cite{Kamenik:2017ghi}, as well as high-$p_T$ mono-tau~\cite{Greljo:2018tzh} or tau-pair production~\cite{Faroughy:2016osc} at the LHC.
Finally, following the proposal in Ref.~\cite{1702.07356}, the BaBar collaboration has recently measured the lepton flavour universality (LFU) ratio in $\Upsilon(3S)$ decays~\cite{2005.01230}
\beq
R_{\tau/\mu}^{\Upsilon(3S)} \equiv \frac{\mathcal B (\Upsilon(3S) \to \tau^+ \tau^-)}{\mathcal B (\Upsilon(3S) \to \mu^+ \mu^-)} = 0.966 \pm 0.008 \pm 0.014\,,
\eeq
where the first (second) uncertainty estimate is due to statistics (systematics). The measured value is within $1.8\sigma$ of the SM prediction $[R_{\tau/\mu}^{\Upsilon(3S)}]_{\rm SM}  = 0.9948(1)$~\cite{1702.07356}. This measurement is probing the $R_{D^{(*)}}$ LFU anomaly through $b\bar b \to \ell^+ \ell^-$ transitions.

In the following we propose a related but potentially more direct test through inclusive di-leptonic $\Upsilon (4S)$
decays by defining
\beq
\mathcal B_{\ell \ell'}^{\Upsilon(4S)} \equiv \mathcal B (\Upsilon(4S) \to \ell^+  \ell^{\prime -} X)\,,
\eeq
as the inclusive dileptonic branching fraction for $\Upsilon(4S)$ decays to a pair of opposite charged leptons of different flavours, where $X$ denotes all the other (hadronic) activity and missing momentum in the event.

This fully inclusive measurement exploits several key capabilities of the Belle II experiment as well as some specific features of $\Upsilon(4S)$ and $b$-hadrons decays. On the experimental side, the excellent beam energy calibration of Super KEK-B can ensure that the $\Upsilon (4S)$  resonance is produced on shell even if its invariant mass is not reconstructed explicitly from the final state. This also allows the non-resonant background to be well estimated from sideband measurements.
On the theory side, this inclusive decay is almost entirely saturated by decays into $B \bar B$ final states.
Moreover, one can analyse the production of the leptons either from an initial $b$-quark decay or from subsequent parts of the decay chain in detail. All in all, ratios of the form
\begin{equation}
R_{\ell\ell'}^{\Upsilon(4S)} \equiv \frac{\mathcal B^{\Upsilon(4S)}_{\ell''\ell}}{\mathcal B^{\Upsilon(4S)}_{\ell''\ell'}}\,,
\end{equation}
where $\ell,\ell',\ell''$ are three different flavours of leptons $e,\mu,\tau$, provide a very interesting ground to probe lepton flavour universality with an inclusive measurement at Belle II\footnote{In principle, a similar test could be envisioned using $\psi(3770)$ at BESIII since $ \mathcal B (\psi(3770) \to D \bar D)\sim 93\%$ \cite{10.1093/ptep/ptaa104}.}, complementary to exclusive measurements accessible to both Belle II and LHCb experiments. In particular, under suitable experimental conditions one can relate
\beq
R_{\tau_{} \ell_{}}^{\Upsilon(4S)}  =  R(X)_{\tau\ell} + \ldots
\eeq
where $\ell=e,\mu$ and $R(X)_{\tau\ell} \equiv \Gamma (B\to X \tau \nu ) / \Gamma (B \to X \ell \nu)$ is the inclusive $B$ decay LFU ratio, which can be precisely computed in the SM as $\mathcal R(X)_{\tau\ell}= 0.223(4)$~\cite{Freytsis:2015qca}. The dots denote corrections due to neutral $B$ meson mixing effects and charm pollution. In the following we discuss both effects and estimate the accuracy with which this ratio can be measured and compared with the SM expectation, in order to extract potential violations of LFU.

\section{Analysis of the decay chain}

\subsection{\texorpdfstring{$\Upsilon(4S)$}{Y(4S)} decay}\label{sec:upsilon4sdecay}

The  $\Upsilon(4S)$ resonance overwhelmingly decays into $B \bar B$ final states. In particular, there is an experimental bound $\mathcal B (\Upsilon(4S) \to B\bar B) >0.96$~\cite{10.1093/ptep/ptaa104}\, but $\mathcal B (\Upsilon(4S) \to B\bar B)$ could actually be even much closer to one. Indeed the dominant non-$B\bar B$ final states are expected to consist in light hadrons mediated by $\Upsilon(4S) \to 3g$ decays as well as decays to lighter bottomonium  states, in particular, $\Upsilon(4S) \to (\Upsilon (n'S),h_b(n''P)) (\pi \pi,\eta, \eta')$ with $n^{\prime,\prime\prime} < 4 $.
Experimental measurements already exist for the latter contributions, in particular $\mathcal B(\Upsilon(4S) \to \Upsilon(1S) + X) < 4 \times 10^{-3} $, $\mathcal B(\Upsilon(4S) \to \Upsilon(2S) \pi^+\pi^-) = 8.2(8) \times 10^{-5} $, $\mathcal B(\Upsilon(4S) \to h_b(1P) \eta) = 2.18(21) \times 10^{-3} $ with other known modes below the $10^{-4}$ level~\cite{Zyla:2020zbs}. In total we thus estimate   $\mathcal B(\Upsilon(4S) \to {\rm bottomonia}) < 7 \times 10^{-3} $.
On the other hand, the $\Upsilon(4S) \to 3g$ decay width can be estimated in NRQCD~\cite{Bodwin:1994jh}. At LO in velocity and QCD expansion, it is given by~\cite{Sang:2020zdv}
\beq
\Gamma(\Upsilon \to 3 g) = 0.0716 \frac{\alpha_s^3 \langle O_1 \rangle_\Upsilon }{m_b^2}\,.
\eeq
Both leading velocity and QCD corrections are of negative sign and thus serve to decrease the above LO estimate~\cite{Sang:2020zdv}. We thus take it as a conservative upper bound. Using $\alpha_s = 0.22$, $m_b = 4.6$ and the upper estimate on $\langle O_1 \rangle_{\Upsilon(4S)} \lesssim \langle O_1 \rangle_{\Upsilon(3S)} = 1.279 $\,GeV${}^3$~\cite{Chung:2010vz} we obtain $\mathcal B(\Upsilon(4S) \to 3 g) \lesssim 2 \times 10^{-3}$\,.
In total we thus estimate that $\mathcal B(\Upsilon(4S) \to B\bar B) \gtrsim 0.99$\,. 

\subsection{Lepton production}\label{sec:leptonproduction}

Having established that $\Upsilon(4S)$ decays almost only into pairs of $B$ mesons, we consider their subsequent decays inclusively, focusing on final states containing leptons $\ell^{(\prime)} = e, \mu, \tau$ in the final state. We can  differentiate between several measurable inclusive dilepton signatures such as
\beq
\Upsilon(4S) \to e^\pm \mu^\mp X\,, \quad \Upsilon(4S) \to \mu^\pm \tau_{\rm had}^\mp X\,, \quad \Upsilon(4S) \to e^\pm \tau_{\rm had}^\mp X\,,
\eeq
where $\tau_{\rm had}$ denotes a $\tau$ lepton reconstructed from its hadronic decays (e.g. $\tau \to 3\pi \nu$)\footnote{These hadronic tau lepton decays need to be efficiently disentangled from backgrounds like hadronic $B$ decays involving three or more charged pions.}.
We will thus define
\beq
R_{\tau_{\rm had} e}^{\Upsilon(4S)} \equiv \frac{\mathcal B^{\Upsilon(4S)}_{\mu\tau_{\rm had}}}{\mathcal B^{\Upsilon(4S)}_{\mu e}} \qquad {\rm and} \qquad R_{\tau_{\rm had} \mu}^{\Upsilon(4S)} \equiv \frac{\mathcal B^{\Upsilon(4S)}_{e\tau_{\rm had}}}{\mathcal B^{\Upsilon(4S)}_{\mu e}}\,.
\eeq
To relate these ratios to inclusive $B$-decay LFU ratios, we need to isolate contributions where each of the two leptons is produced in a separate $B$-meson decay and suppress backgrounds where one or both leptons do not originate from a direct semileptonic $B$ decay. Requiring different opposite-sign lepton flavour final states removes such contributions from $\Upsilon (4S) \to X + ((\bar b b) \to \ell^+\ell^-) $,  $b \to q ((c \bar c) \to \ell^+\ell^-) $ as well as from rare FCNC (semileptonic) $B$ and charm decays.\footnote{The exceptions with $\ell^\pm=\tau^\pm$ where one of the taus decays leptonically and the other hadronically, leading to a final state with a hadronic tau and a lighter lepton, turn out to be numerically negligible as they are suppressed by small $\Upsilon (4S) \to X + $bottomonium, $B \to X +$charmonium~\cite{Zyla:2020zbs}, and $B \to X \tau^+ \tau^-$~\cite{Kamenik:2017ghi}  branching ratios, respectively.}
This approach is however not effective against contamination from $b \to q (c\to q' \ell^+ \nu)({\bar c} \to q'' \ell^{\prime -} \nu)$ and $b \to (c \to q \ell^+ \nu) \ell^{\prime -} \nu)$ transitions, which we will address in Sec.~\ref{sec:sameBmeson}.

For the moment we assume that each of the two different lepton tags originates from a separate $B$-meson decay pattern. We will focus on $R_{\tau_{\rm had} \mu}^{\Upsilon(4S)}$ for the time being, but a very similar analysis can be performed for $R_{\tau_{\rm had} e}^{\Upsilon(4S)}$ swapping muons and electrons in the discussion.
 The single hadronic tau can be produced in the quark-level transition chains
\begin{align}
b &\to q \tau \nu\,, \nonumber\\
b & \to q \bar q'  (c \to q^{\prime\prime} \tau \nu )\,.
\end{align}
On the other hand a single muon (or equivalently electron) can originate from
\begin{align}
b &\to q \mu \nu\,, \nonumber\\
b & \to q \bar q'  (c \to q^{\prime\prime} \mu \nu )\,,\nonumber\\
b &\to q (\tau \to \mu \nu \nu) \nu\,, \nonumber\\
b & \to q \bar q'  (c \to q^{\prime\prime} (\tau \to \mu \nu \nu) \nu )\,.
\end{align}
Inclusive semileptonic $b$-hadron decays (i.e. $b\to q \ell \nu$) are well under theoretical control and thus the associated rates can be well predicted, including possible effects of LFU violation~\cite{Freytsis:2015qca}. The same cannot necessarily be said for inclusive semileptonic charm decays~\cite{Gambino:2010jz}\,, which thus represent a challenging background. One could imagine that the charge of the leptons could help us to disentangle the origin of the lepton, either from a $b$ or from a $c$-quark. However, one should take into account that in approximately half of the cases, the $\Upsilon(4S)$ decays into neutral $B$ mesons, which can oscillate and spoil the identification between the charge of initial quark and that of the lepton. We discuss strategies how to mitigate this effect next.

\subsection{Mixing effects}\label{sec:mixingeffects}

We first define the amplitudes
$
A(B^0\to \ell^-X), A(B^0\to \ell^+ X), A(\bar{B}^0\to \ell^+ X), A(\bar{B}^0\to \ell^- X)
$
embedding the complete meson decay chains (for instance it may contain $B\to D\pi$ followed by $D\to \ell X$), so that the lepton is not necessarily produced by the decay of the $b$-quark. However, it is not produced by the decay of the light quark in the $B$, which means that in the isospin limit, we have equalities of the type:
\begin{equation}
A(B^0\to \ell^-X) = A(B^+\to \ell^-X)=A_{\ell^-} \qquad A(\bar{B}^0\to \ell^-X) = A(B^-\to \ell^-X)=\bar{A}_{\ell^-} \qquad \ldots
\end{equation}
where the presence/absence of the bar indicates the charge of the $b$-quark inside the $B$-meson and the subscript denotes the charge and flavour of the lepton.

If we look for $\Upsilon(4S)\to \ell_1 \ell_2 X$ (with 1 and 2 being different, either by flavour or charge) through
an intermediate $B^0\bar{B}^0$ state,  we can use the description introduced for the study of CP violation from the production of an intricated $B$-meson pair (sec 1.2.3 in Ref.~\cite{Harrison:1998yr}), leading to the time-dependent rate where one of the two $B$ mesons decay into a state containing $\ell_1$ at a time $t_1$ and the other one into a state containing $\ell_2$ at a time $t_2$, leading to
\begin{equation}
R(t_1,t_2)=
  Ce^{-\Gamma(t_1+t_2)}
   \Bigg[{\mathcal I}
          -\cos(\Delta m(t_1-t_2)){\mathcal C}      +2\sin(\Delta m(t_1-t_2)){\mathcal S}\Bigg]\,,
          \end{equation}
          where $C$ is a normalisation coming from angular integration, $\Delta m$ is the difference of mass between the two mass eigenstates, $\Gamma$ is their average width, the approximations $|q/p|=1$ and $\Delta \Gamma=0$ have been used,
            and we have
\begin{eqnarray}
{\mathcal I}&=&\left[(|A_1|^2+|\bar{A}_1|^2)(|A_2|^2+|\bar{A}_2|^2)-4{\rm Re}\left(\frac{q}{p}A_1^*\bar{A}_1\right)
                 {\rm Re}\left(\frac{q}{p}A_2^*\bar{A}_2\right)\right]\,,\\
{\mathcal C}&=&\left[(|A_1|^2-|\bar{A}_1|^2)(|A_2|^2-|\bar{A}_2|^2)
           +4{\rm Im}\left(\frac{q}{p}A_1^*\bar{A}_1\right)
                 {\rm Im}\left(\frac{q}{p}A_2^*\bar{A}_2\right)\right]\,,\\
{\mathcal S}&=&\left[{\rm Im}\left(\frac{q}{p}A_1^*\bar{A}_1\right)(|A_2|^2-|\bar{A}_2|^2)
           +(|A_1|^2-|\bar{A}_1|^2)
                 {\rm Im}\left(\frac{q}{p}A_2^*\bar{A}_2\right)\right]\,.
\end{eqnarray}

In order to prevent too large effects from mixing,
one could consider cutting too large time differences $|t_1-t_2| $, so that there has not been enough time for the evolution to take place.
Cutting at $|t_1-t_2|={\alpha}/{\Delta m}$ (where $\alpha$ is the cut parameter) leads to
\begin{equation}
        R^\alpha_{B^0\bar{B^0}}\equiv \int_{-\alpha/\Delta m}^{\alpha/\Delta m}R(t')\mathrm{d}t'=  \frac{2C}{\Gamma^2}
   \Bigg[(1-e^{-\frac{\alpha}{x}}){\mathcal I}
       -\frac{1-e^{-\frac{\alpha}{x}}(\cos\alpha-x\sin\alpha)}{1+x^2}{\mathcal C}\Bigg]\,,
\end{equation}
where $x=\Delta m/\Gamma\simeq 0.769$~\cite{Amhis:2019ckw}. In the case of $B^+B^-$, where no mixing is involved, we have
\begin{equation}
        R^\alpha_{B^+B^-}\equiv \int_{-\alpha/\Delta m}^{\alpha/\Delta m}R(t')\Big|_{\Delta m=0}\mathrm{d}t'=  \frac{2C}{\Gamma^2}
   (1-e^{-\frac{\alpha}{x}})({\mathcal I}
       -{\mathcal C})\,.
\end{equation}
Denoting the result without cut in the time difference as $R_{BB}\equiv R_{BB}^\infty$, we see then that the effect of mixing corresponds to
\begin{equation}
        R^\alpha_{B^0\bar{B^0}}-R^\alpha_{B^+B^-}=
  \left(1-\frac{e^{-\frac{\alpha}{x}}(1+x^2-\cos\alpha+x\sin\alpha)}{x^2}\right)(R_{B^0\bar{B^0}}-R_{B^+B^-})\,,
\end{equation}
whereas we have
\begin{equation}
        R^\alpha_{B^+B^-}=(1-e^{-\frac{\alpha}{x}}) R_{B^+B^-}\,,
\end{equation}
\begin{figure}
    \centering
    \includegraphics[width=0.6\textwidth]{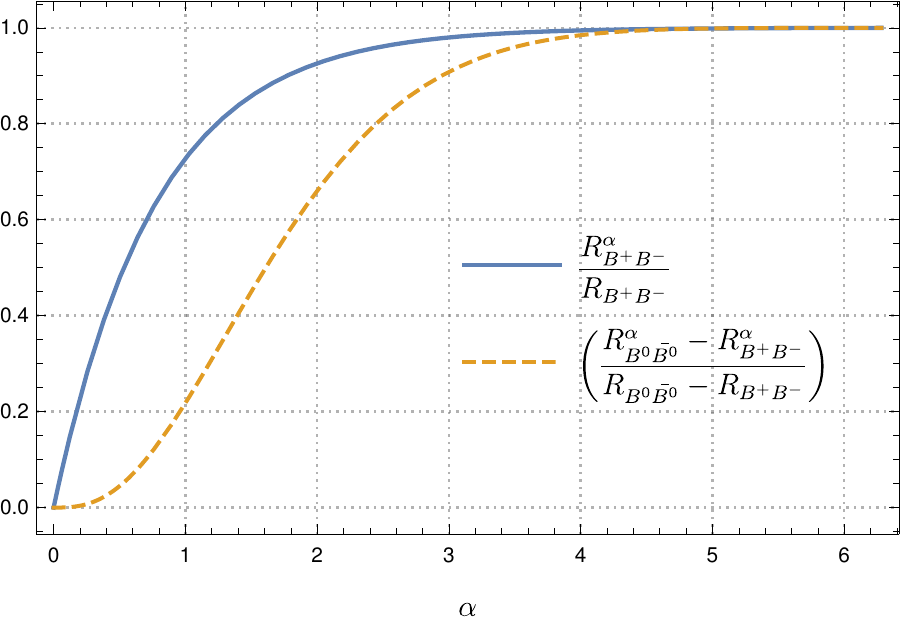}
    \caption{Relative branching fraction (blue solid) and mixing effect (yellow dashed) as a function of the cut parameter $\alpha$.}
    \label{fig:Relative_mixing_effect_alpha}
\end{figure}
In \cref{fig:Relative_mixing_effect_alpha}, we illustrate the impact of the cut on the branching fraction of the decay into $B^+B^-$ and on the mixing effect. We see that this cut can efficiently suppress the impact of mixing while keeping a large fraction of the $B^+B^-$ signal.

We can study the impact of this cut on the $\Upsilon(4S)$ decay rate. Introducing $\rho={\cal B}(\Upsilon(4S)\to B^+B^-)=0.514\pm 0.006$~\cite{Amhis:2019ckw}, we have
${\cal B}(\Upsilon(4S)\to B^0\bar{B}^0)=0.486\pm 0.006=1-\rho-\epsilon$ with $\epsilon<0.01$ according to our estimates. We have then the total rate $R={\cal B}(\Upsilon(4S)\to BB)$
\begin{equation}
R=\rho R_{B^+B^-}+(1-\rho-\epsilon)R_{B^0\bar{B}^0}
  =(1-\epsilon) R_{B^+B^-}+(1-\rho-\epsilon) (R_{B^0\bar{B}^0}-R_{B^+B^-})\,,
\end{equation}
where the first term corresponds to the rate without mixing (in the isospin limit) and the second term to the contamination due to mixing. Cutting $|t_1-t_2|\leq \alpha/\Delta m$, we have
\begin{eqnarray}
 R^\alpha
& =&  (1-\epsilon) R_{B^+B^-}^\alpha+(1-\rho-\epsilon)(R_{B^0\bar{B}^0}^\alpha-R_{B^+B^-}^\alpha)\\
&=& (1-\epsilon) (1-e^{-\frac{\alpha}{x}}) R_{B^+B^-}
    +(1-\rho-\epsilon)\left(1-\frac{e^{-\frac{\alpha}{x}}(1+x^2-\cos\alpha+x\sin\alpha)}{x^2}\right) (R_{B^0\bar{B}^0}-R_{B^+B^-})\,.\label{eq:Ralpha}
\end{eqnarray}
The first term in Eq.~\eqref{eq:Ralpha} goes like $O(\alpha)$ whereas the second term goes like $O(\alpha^3)$. Moreover, $(R_{B^0\bar{B}^0}-R_{B^+B^-})$ is equal to
\begin{equation}
R_{B^0\bar{B^0}}-R_{B^+B^-}
 = \frac{2C}{\Gamma^2}\frac{x^2}{1+x^2}
  {\mathcal C}\,,
\end{equation}
so it goes like $O(x^2)$ and it is isospin suppressed. Then the second term in Eq.~\eqref{eq:Ralpha}, corresponding to the mixing effects, is suppressed significantly.

From \cref{fig:Relative_mixing_effect_alpha}, we see that $\alpha=0.53$ would ensure that the second contribution is $O(1\%)$ of $R_{B^+B^-}$, taking into account the suppressions by the $\alpha$-dependent factor, by $x^2/(1+x^2)$ and by $1-\rho$ (but not taking into account the isospin suppression, which would further suppress this term). On the other hand, the first contribution in Eq.~(\ref{eq:Ralpha}) would be half of $R_{B^+B^-}$ (essentially $R$ without the effect of mixing in the isospin limit).
More generally, a fit to $R^{\alpha}$ as a function of $\alpha$ would allow one to put a bound on  $(R_{B^0\bar{B}^0}-R_{B^+B^-})$ and to extract $R_{B^+B^-}$ directly.

\subsection{Charm pollution}\label{sec:charmpollution}

As shown in the previous section, cutting on the time difference of the two decaying $B$ mesons can suppress mixing effects and allow one to distinguish leptons originating from $B$ and charm decays by charge. However, we still need to quantify the expected initial amount of charm contamination.
Under the assumption that each tagged lepton originates from a separate $B$ decay chain (which we will relax in the next section), the ratio $R_{\tau_{\rm had} \mu}^{\Upsilon(4S)}$ can be conveniently expressed in terms of
\beq
[R_{\tau_{\rm had} \mu}^{\Upsilon(4S)}]^{-1}  = \frac{\mathcal B (B \to  X \mu \nu )  + \mathcal B (B \to X (h_c \to X' \mu \nu)) }{\mathcal B (B \to X \tau_{\rm had}  \nu) + \mathcal B (B \to X (h_c \to  X' \tau_{\rm had} \nu))}   + \frac{\mathcal B (\tau \to \mu \nu \bar \nu)}{\mathcal B(\tau \to \tau_{\rm had})}\,,
\eeq
where $h_c$ denotes any weakly decaying charmed hadron, i.e. $D^+$, $D^0$, $D_s$, $\Lambda_c$ and their charge conjugates. As discussed above, using charge ID, but also possibly a cut on leptons not originating from the secondary vertex (i.e. from $B$ decays), it should be possible to suppress contributions where the leptons originate from secondary charm or, in the case of muons, tau decays, by efficiency factors $\epsilon^{(i)} \ll 1$. This allows us to simplify the above expression and write it in terms of the inverse of the inclusive ratio $R(X)_{\tau\mu}$. We obtain
\begin{align}
[R_{\tau_{\rm had} \mu_{b}}^{\Upsilon(4S)}]^{-1}  = &
\Bigg\{
 [R(X)_{\tau\mu}]^{-1} \left[ 1 -  \epsilon^{(3)}_{\rm } \frac{ \mathcal B (B \to X (h_c \to X' \tau \nu)) }{\mathcal B (B \to  X \tau_{\rm } \nu )} \right]    \nonumber \\ &
\qquad + \epsilon^{(1)}_{\rm } {\mathcal B (\tau \to \mu \nu \bar \nu)} + \epsilon^{(2)}_{\rm } \frac{\mathcal B (B \to X (h_c \to X' \mu \nu)) }{\mathcal B(B \to  X \tau_{\rm } \nu) )} \Bigg\} [\mathcal B (\tau \to \tau_{\rm had})]^{-1}\,,
 \label{eq:R2}
\end{align}
where  $\mu_{b}$ denotes muons consistent with originating from the secondary (i.e. $b$-decay) vertex and we have already used the fact that  $\mathcal B (B \to X \tau  \nu) \gg \mathcal B (B \to X (h_c \to X' \tau \nu))$\,, which we verify below.

We can estimate the size of all three corrections on the right-hand side of Eq.~\eqref{eq:R2}  (up to the $\epsilon^{(i)}$ efficiencies) based almost purely on experimental information. Starting with the $\epsilon^{(1)}$ term, $\mathcal B(\tau \to \mu \bar \nu \nu)= (17.39 \pm 0.04) \%$~\cite{Zyla:2020zbs} we see, that even without cuts (for $\epsilon^{(1)} \simeq 1$)  it leads to an order $4\%$ (computable) systematic effect in $\mathcal R(X)_{\tau\mu}$.

We estimate the second and third term thanks to the identity
\beq
\mathcal B (B \to X (h_c \to X' \ell \nu)) = \mathcal B (B\to X_c ) \sum_i f(c \to h_c^{(i)}) \mathcal B (h_c^{(i)} \to X \ell \nu)\,,
\eeq
where the sum runs over all weakly decaying charmed hadrons and $f(c\to h_c^{(i)})$ are the corresponding fragmentation functions. We use Ref.~\cite{Aubert:2006mp} for the charm-inclusive decay branching ratio $\mathcal B (B \to X_c) = (97\pm4) \%$  and Ref.~\cite{Lisovyi:2015uqa} for the charm fragmentation functions. Note that the above estimate relies on factorization of the inclusive $B$-decay amplitudes and is thus subject to related theoretical uncertainties. In addition, the application of charm fragmentation functions extracted from high energy $e^+ e^-$ and $ep$ collision data to $B$ decays carries further systematic errors. Consequently, our background evaluations should be taken as order-of-magnitude estimates, which are however sufficient for our purpose.
For ${\mathcal B} (D_s \to X \mu \nu)$ we use values measured by CLEO for the electron in the final state \cite{Asner:2009pu}  ${\mathcal B} (D_s \to X e \nu)= (6.52\pm 0.39\pm 0.15)\%$ which can serve as an effective upper bound on  ${\mathcal B} (D_s \to X \mu \nu)$ assuming $e-\mu$ LFU in charm decays.
 We also use $\mathcal B ( D^+\to X e\nu) = 0.1607 \pm 0.0030$ and  $\mathcal B ( D^0\to X e\nu) = 0.0649\pm 0.0011$ ~\cite{Zyla:2020zbs}. Finally, we obtain
\begin{eqnarray}
\mathcal B (B \to X (h_c \to X' \mu \nu)) &=& \mathcal  B(b \to X_c) \{ f (c\to D^0)
\mathcal  B(D^0 \to X \mu\nu)
+ f (c\to D^+)
 \mathcal  B(D^+ \to  X \mu\nu)  \nonumber\\
&& + f (c\to D_s)
\mathcal  B(D_s \to  X \mu \nu) + f (c\to \Lambda_c) \mathcal  B(\Lambda_c \to  X \mu\nu)   + \ldots\}  \lesssim 0.088\,,\nonumber\\
\mathcal B (B \to X (h_c \to X' \tau \nu)) &=& \mathcal  B(b \to X_c) \{ f (c\to D^+)
 \mathcal  B(D^+ \to \tau \nu)  + f (c\to D_s)
\mathcal  B(D_s \to \tau \nu)    + \ldots\}
\simeq 0.0067\,.
    \end{eqnarray}
These values are to be compared with the LEP experimental determination of $\mathcal B(b\to q\tau \nu)\simeq \mathcal B(B\to X \tau \nu)= (2.41 \pm 0.23) \%$~\cite{Freytsis:2015qca}.  In particular, before cuts and without lepton charge ID (for $\epsilon^{(2)} \simeq 1$) the second term  in Eq.~\eqref{eq:R2} would represent a dominant $80\%$ systematic effect in the determination of  $\mathcal R(X)_{\tau\mu}$.
Finally, the effect of the $\epsilon^{(3)}$ term
before cuts (for $\epsilon^{(3)} \simeq 1$) represents a relative $28\%$ systematic effect on the determination of  $\mathcal R(X)_{\tau\mu}$.

In summary, the term with $\epsilon^{(1)}$ is small thanks to the low value of $\mathcal B(\tau \to \mu \bar \nu \nu)$, whereas the factors of $\epsilon^{(2)}$ and $\epsilon^{(3)}$ have large values but are related to charm pollution, which (hopefully) can be reduced thanks to charge ID leading to small efficiencies $\epsilon^{(2,3)}$.

\subsection{Leptons emitted from the same \texorpdfstring{$B$}{B}-meson} \label{sec:sameBmeson}

Lastly we need to consider backgrounds where both leptons are of different charge and flavour, but originate from the same $B$-decay chain, corresponding to the parton-level chain
\beq
b \to q (c\to q' \ell^+ \nu)(\bar c \to q'' \ell^{\prime -} \nu) \quad {\rm and} \quad b \to (c \to q \ell^+ \nu)  \ell^{\prime -} \nu\,.
\eeq
Denoting these processes collectively as $B \to X \ell \ell'$, and assuming they can be suppressed by cutting on leptons not originating from the secondary vertex (i.e. from $b$ decays), we can again write the relative correction to Eq.~\eqref{eq:R2} due to these contributions expanded to leading order in all $\epsilon^{(i)}$ as
\beq
[R_{\tau_{\rm had} \mu_{b}}^{\Upsilon(4S)}]^{-1}  =  [{ R(X)_{\tau\mu} \mathcal B (\tau \to \tau_{\rm had})}]^{-1} \left[ 1 - \epsilon^{(4)} \frac{\mathcal B (B\to X_{\rm }) \mathcal B (\bar B \to X \tau e)}{\mathcal B( B \to X \tau \nu) \mathcal B( \bar B \to X e \nu)} + \epsilon^{(5)} \frac{\mathcal B (B\to X_{\rm }) \mathcal B (\bar B \to X \mu e)}{\mathcal B( B \to X \mu \nu) \mathcal B( \bar B \to X e \nu)}  \right] + \ldots\,
\eeq
where the inclusive hadronic $B$-decay branching ratio is denoted as $\mathcal B (B \to X_{\rm}) \lesssim 1- \sum_\ell {\mathcal B (B \to X \ell \nu)} \simeq 0.76$, we take $\mathcal B (B \to X_c e \nu) \simeq  \mathcal B (B \to X_c \mu \nu) \simeq 0.11$~\cite{Zyla:2020zbs}, and the ellipsis denotes the remaining corrections on the right-hand side of Eq.~\eqref{eq:R2}. Using the numerical values given above we  obtain for the relevant $b \to (c \to q \ell^+ \nu)  \ell^{\prime -} \nu$ transitions
\begin{eqnarray}
\mathcal B(B\to  X(h_c \to X' e \nu) \tau \nu)  &=& \mathcal B(B\to X_c\tau \nu) [ f (c\to D^0) \mathcal  B(D^0 \to X e\nu)
+ f (c\to D^+) \mathcal  B(D^+ \to  X e\nu)  \nonumber\\
&& + f (c\to D_s) \mathcal  B(D_s \to  X e \nu) + f (c\to \Lambda_c) \mathcal  B(\Lambda_c \to  X e\nu)   + \ldots ]  \simeq 0.0021\,,\nonumber\\
\mathcal B (B\to X(h_c \to X' \tau \nu) e \nu) &=&\mathcal B (B \to X_c e \nu) f (c\to D_s) \mathcal  B(D_s \to \tau \nu) \simeq 0.00042\,, \nonumber\\
\mathcal B(B\to  X(h_c \to X' e \nu) \mu \nu)  &=& \mathcal B(B\to X_c\mu \nu) [ f (c\to D^0) \mathcal  B(D^0 \to X e\nu)
+ f (c\to D^+) \mathcal  B(D^+ \to  X e\nu)  \nonumber\\
&& + f (c\to D_s) \mathcal  B(D_s \to  X e \nu) + f (c\to \Lambda_c) \mathcal  B(\Lambda_c \to  X e\nu)   + \ldots ]  \simeq \, 0.0095\nonumber\\
\mathcal B (B\to X(h_c \to X' \mu \nu) e \nu) &=&\mathcal B (B \to X_c e \nu)  [ f (c\to D^0) \mathcal  B(D^0 \to X \mu\nu)
+ f (c\to D^+) \mathcal  B(D^+ \to  X \mu\nu) \nonumber\\
&& + f (c\to D_s) \mathcal  B(D_s \to  X \mu \nu) + f (c\to \Lambda_c) \mathcal  B(\Lambda_c \to  X \mu\nu)   + \ldots ]  \lesssim \, 0.010.
\label{eq26}
\end{eqnarray}

Finally, for the decay chain $ b\to q {c \bar c} (c \to q \ell \nu ) (\bar c \to q^\prime \ell' \nu)$, using $\mathcal B(B \to X_{c \bar c}) \simeq 22 \% $~\cite{Abbaneo:2001bv}
 and after including $c \to q \ell \nu $ and $\bar c \to q^\prime \ell' \nu$ transition rates, we find
\begin{align}
\mathcal B(B \to X  (h_c \to X' e \nu ) (h_{\bar c} \to X'' \tau \nu) ) = &  \mathcal B(B \to X_{c \bar c} ) f (\bar c\to \bar D_s)\mathcal B (\bar D_s\to \tau \nu) \nonumber\\
 & \times \left[ f (c\to D^0) \mathcal  B(D^0 \to X e\nu) + f (c\to D^+)  \mathcal  B(D^+ \to  X e\nu)  \right. \nonumber\\
& {\phantom{\times [\,}} \left. + f (c\to D_s)  \mathcal  B(D_s \to  X e \nu) + f (c\to \Lambda_c) \mathcal  B(\Lambda_c \to  X e\nu)  \right] \simeq 0.0001\,,\nonumber\\
\mathcal B(B \to X  (h_c \to X' e \nu ) ( h_{\bar c} \to X'' \mu \nu) ) = &  \mathcal B(B \to X_{c \bar c} ) \nonumber\\
 & \times \left[ f (c\to D^0) \mathcal  B(D^0 \to X \mu\nu) + f (c\to D^+)  \mathcal  B(D^+ \to  X \mu\nu)  \right. \nonumber\\
& {\phantom{\times [\,}} \left. + f (c\to D_s)  \mathcal  B(D_s \to  X \mu \nu) + f (c\to \Lambda_c) \mathcal  B(\Lambda_c \to  X \mu \nu)  \right] \,\nonumber\\
 & \times \left[ f (c\to D^0) \mathcal  B(D^0 \to X e\nu) + f (c\to D^+)  \mathcal  B(D^+ \to  X e\nu)  \right. \nonumber\\
& {\phantom{\times [\,}} \left. + f (c\to D_s)  \mathcal  B(D_s \to  X e \nu) + f (c\to \Lambda_c) \mathcal  B(\Lambda_c \to  X e\nu)  \right] \simeq 0.0018\,.\end{align}

Putting these values together we observe that these backgrounds are individually comparable in size to the signal (i.e. they would represent approximately $80\%$ and $150\%$ relative corrections, respectively)  in absence of cuts to suppress them (for $\epsilon^{(4,5)} \simeq 1$).
While they are similar in magnitude, they are highly correlated and contribute with opposite signs, so
that they tend to cancel to a degree for $\epsilon^{(4)} \simeq \epsilon^{(5)}$. In fact, the two terms become exactly equal in the limit where one can neglect charm decays to muons and taus (in the ratio $R_{\tau e}^{\Upsilon(4S)}$ these would be charm decays to electrons and taus).
On the other hand, contrary to the corrections outlined in Eq.~\eqref{eq:R2}, the corrections considered in this section cannot be suppressed using only lepton charge ID.  This highlights the crucial importance of discriminating against leptons originating from the same $B$-decay chain, for instance through geometrical considerations. An alternative strategy could consist in discriminating leptons arising from the secondary ($B$-decay) vertices from those arising further down in the decay chains. A quantitative assessment of the feasibility of either of the two approaches through an appropriate experimental analysis would require a dedicated experimental study and is beyond the scope of this work.

\section{Conclusions}

Relying on the specific properties of B-factories and in particular the Belle II experiment, we have proposed to compare the inclusive rates of
$\Upsilon(4S) \to e^\pm \mu^\mp X$,  $\Upsilon(4S) \to \mu^\pm \tau_{\rm had}^\mp X$ and $\Upsilon(4S) \to e^\pm \tau_{\rm had}^\mp X$.
This measurement can be related to the ratio $R(X)_{\tau\ell} \equiv \Gamma (b\to X \tau \nu ) / \Gamma (b \to X \ell \nu)$ ($\ell=e$ or $\mu$), once appropriate experimental cuts are applied to suppress the effects of neutral $B$ mixing and leptons emitted from rare FCNC (semileptonic) $B$ decays, as well as secondary charmonium, charm and tau decays.  The feasibility of our proposal crucially assumes that hadronically decaying tau leptons originating from the $B$ decay vertices can be efficiently disentangled from backgrounds (e.g. from hadronic $B$ decays involving three or more charged pions) at Belle II. A dedicated experimental study of this is however beyond the scope of the present paper.

We have focused on the case of $R_{\tau \mu}^{\Upsilon(4S)} \simeq R(X)_{\tau\mu}$, but our discussion applies equally well to the tau-electron combination, swapping the roles played by electrons and muons.
The current deviations in $B\to D^*\ell\nu$ and $B\to D\ell\nu$ when $\tau$ channels are compared to electronic or muonic modes are at the level of 10\% (for the LFU ratios of branching ratios) and provide a benchmark for the target sensitivity of our proposal. This is illustrated by the very simple case where NP mimics the $V-A$ structure of $b \to c \tau  \nu$ currents in the SM, leading to a universal rescaling of all $b\to c\tau\nu$ branching ratios.

Given our estimates, the systematic uncertainties in the determination of $R(X)_{\tau\ell}$ from a measurement of $R_{\tau \ell}^{\Upsilon(4S)}$ could be brought below a given value ($\epsilon_{\rm sys}$) provided that (1) cuts on the $B-\bar B$ impact parameter difference can suppress the neutral $B$ meson mixing effects below $\epsilon_{\rm sys}$ combined with an efficient lepton charge ID to suppress semileptonic charm-decay contamination; (2a) multiple leptons originating from the same $B$ decay chain can be suppressed to better than $\epsilon_{\rm sys}$ or alternatively (2b) leptons arising from the secondary ($B$-decay) vertices can be discriminated against those arising further down in the decay chains to roughly better than $\epsilon_{\rm sys}$. Further dedicated experimental studies are needed to establish the actually attainable precision by Belle II.

In summary, we have proposed a novel method to test the persistent hints of violation of LFU observed in semileptonic $B$ decays. This measurement would constitute an additional and potentially competitive probe of LFU violations in $b\to c\ell\nu$ transitions, complementary to exclusive measurements and accessible in the Belle II environment.

\begin{acknowledgments}
We would like to thank Bo\v{s}tjan Golob for useful discussions and comments.
JFK and SF acknowledge the financial support from the Slovenian Research Agency
(research core funding No. P1-0035). This project has received support from the European Union’s Horizon 2020 research and innovation programme under the Marie Skodowska-Curie grant agreement No 860881-HIDDeN.
\end{acknowledgments}

\bibliographystyle{elsarticle-num}
\bibliography{ref}

\end{document}